\documentclass[aps,prd,draft,preprint,showpacs,eqsecnum]{revtex4}

\def\vX{{\bf X}}

\begin{document}


\title{General-relativistic perturbation equations
for the dynamics of elastic deformable astronomical bodies
expanded in terms of generalized spherical harmonics}

\author{Chongming Xu}
 \email{cmxu@njnu.edu.cn}
\author{Xuejun Wu}
 \email{xjwu@njnu.edu.cn}
\affiliation{Department of physics, Nanjing Normal University,
 Nanjing 210097, China}
\author{Michael Soffel}
 \email{soffel@rcs.urz.tu-dresden.de}
\affiliation{Lohrmann Observatory, Technical University Dresden,
D-01062 Dresden, Germany}

\date{\today}


\begin{abstract}
In our previous paper, based on the Carter \& Quintana framework and the
Damour-Soffel-Xu scheme, we deduced a complete and closed set of
post-Newtonian dynamical equations for elastically deformable
astronomical bodies. In this paper, we expand the general
relativistic perturbation equations of elastic deformable bodies
(field equations, stress-strain relation, Euler equation) in terms
of Generalized Spherical Harmonics. This turns the set of
complicated partial differential equations into a set of ordinary
differential equations. This will be useful for numerical
applications that mainly deal with the global dynamics of the
Earth.
\end{abstract}


\maketitle


\section{Introduction}

In a series of papers \cite{xu01,xu03} the dynamical equations for
elastic deformable astronomical bodies in the framework of
Einstein's theory of gravity (GRT) were derived. Fields of
application range from problems of seismology of astronomical
bodies including the problem of normal modes to the global motion
of such bodies in space. Actually one of our main interests is the
problem of global geodynamics where the accuracy of modern
geodetic space techniques such as VLBI has long reached the level
where effects from relativity have to be taken into account.
Unfortunately the field of relativistic global geodynamics has not
yet reached a satisfactory level despite the fact that published
amplitudes of nutation series are of $\mu$as (microarcsec)
accuracies.

This paper presents another necessary step to improve this
situation. It extends our previous work mentioned above that was
based upon two frameworks: (1) the Carter-Quintana formalism \cite
 {cart72,cart73} for the description of elastic deformable bodies
in GRT by means of a displacement field and (2) the
Damour-Soffel-Xu one (the DSX scheme
 \cite{damo91,damo92,damo93,damo94}) on relativistic celestial
mechanics in the first post-Newtonian approximation to GRT.

Here we shall expand the general relativistic perturbation
equations for the dynamics of  elastic deformable bodies in terms
of Generalized Spherical Harmonics (GSH). This turns the set of
partial differential equations for scalars, vectors and tensors
into a set of ordinary differential equations. Actually in the
Newtonian approximation numerical programs that have been written
mainly to deal with the global dynamics of the Earth (e.g., Wahr
 \cite{wahr82}, Dehant \cite{deha97}, Schastok \cite{scha97}) by
considering its elastic properties in a local framework usually
integrate such (Newtonian) equations. In general relativity, the
general expansion  for vectors and tensors  by means of GSH has
been discussed in several papers \cite{thor80,dam91}. Here, we
apply these methods and expand the general relativistic dynamical
equations for elastic deformable, nonrotating astronomical bodies
by means of GSH.

In the following we consider the dynamics of one of these bodies
in its own local coordinate system $(cT,\vX)$ restricting
ourselves to a spherically symmetric non-rotating relaxed ground
state. The extension to some rotating axially symmetric ground
state with nonvanishing dynamical ellipticity so that problems of
precession and nutation can be treated will be the subject of
another paper. In Sec.~II, we briefly review the main results in
Ref.\cite{xu01}. We also rewrite the equations and introduce
suitable notation in order to facilitate the comparison with the
corresponding Newtonian equations (Ref.\cite{wahr82}).

In Sec.~III, Generalized Spherical Harmonics are introduced. Our
main new results are presented in Sec.~IV, where the post
Newtonian perturbation equations (PDE) are expanded so that they
reduce to a set of ordinary differential equations (ODE). In
Sec.~V some conclusion can be found.


\section{perturbed field equation and PN Euler equation}

Let us first recall some relations that will be relevant for this
paper.  In the DSX-formalism, the Einstein field equations to
first post-Newtonian order can be written as \cite{damo91}
\begin{eqnarray}
\nabla ^2 W - \frac{1}{c^2} \frac{\partial ^2 W}{\partial T ^2}
  &=& -4\pi G \Sigma + O(4) \, ,\label{field1} \\
\nabla ^2 W^a &=& - 4\pi G \Sigma ^a + O(2) \, , \label{field2}
\end{eqnarray}
where $W$ and $W^a$ are the scalar and vector potentials that
describe the gravitational interaction, and the gravitational
mass-density and mass-current-density, $\Sigma$ and $\Sigma^a$, are
related with the energy momentum tensor by $\Sigma = ( T^{00} +
T^{aa})/c^2 $, $ \Sigma ^a = T^{0a}/c $. We shall often abbreviate
the order symbol $O(c^{-n})$ simply by $O(n)$.

For a non-rotating static spherically symmetric body (unperturbed
state), the energy-momentum tensor takes the form
\begin{eqnarray}
T ^{00} &=& \rho c ^2 \left( 1 + \frac{2 W }{c^2} \right) +
O(2)\, , \\
T ^{aa} &=& 3 p \left( 1 - \frac{2 W }{c^2} \right) + O(4)\, , \\
T ^{0a} &=& O(3) \, ,
\end{eqnarray}
where $\rho c^2 $ is the energy density and $p$ is the pressure.
The field equations reduce to
\begin{eqnarray}
\nabla ^2 W  &=&  - 4 \pi G  \left[ \rho ^*  \left( 1 +\frac{2W}{c^2} \right)
  + \frac{2p}{c^2} \right] + O(4) \, , \label{field3} \\
\nabla ^2 W^a &=&  O(2) \, , \label{field4}
\end{eqnarray}
where $\rho^* = \rho + \frac{p}{c^2}$ is the chemical potential
per unit volume. For a static equilibrium configuration the scalar
potential $W$ is a function of the radial coordinate $r$ only, and
the gravito-magnetic potential $W^a$ vanishes.

\bigskip
Now we consider a perturbed state. The perturbed field equations
for the Eulerian variations of gravitational potentials, $\delta
W$ and $\delta W ^a$ read
\begin{eqnarray}
\nabla ^2 \delta W - \frac{1}{c^2}
  \frac{\partial ^2 \delta W}{\partial T ^2}
  &=& -4\pi G \delta \Sigma + O(4) \, ,\label{df1} \\
\nabla ^2 \delta W^a &=& - 4\pi G \delta \Sigma ^a + O(2) \, ,
  \label{df2}
\end{eqnarray}
where, by using Eqs.(4.27), (4.31), (4.40) and (4.41) of our
previous paper (Ref.\cite{xu01}),
\begin{eqnarray}
\delta \Sigma &=& \delta \rho + \frac{1}{c^2} \left(
  2 \rho \delta W + 2 W \delta \rho  + 3 \delta p
  \right) \nonumber \\
  &=& - \nabla \cdot (\rho {\bf s}) - \frac{1}{c ^2} \Bigl[
   \nabla \cdot (p {\bf s}) + 3 \rho {\bf s} \cdot \nabla W
  + \rho \delta W + 2 \nabla \cdot (\rho W {\bf s}) + 3\kappa
  \nabla \cdot {\bf s} \Bigr] + O(4) \, , \\
\delta \Sigma ^a &=& \rho \dot{s} ^a + O(2) \,
\end{eqnarray}
and  $\kappa $ is the compression modulus. Then the field
equations can be expressed as
\begin{eqnarray}
\nabla ^2 \delta W - \frac{1}{c^2}
   \frac{ \partial ^2 \delta W}{\partial T ^2}
   &=& 4 \pi G \Biggl\{ \nabla \cdot (\rho {\bf s})
   + \frac{1}{c ^2} \Bigl[
   \nabla \cdot (p {\bf s}) + 3 \rho {\bf s} \cdot \nabla W
   + \rho \delta W  \nonumber \\
  &&  + 2 \nabla \cdot (\rho W {\bf s}) + 3\kappa
   \nabla \cdot {\bf s} \Bigr] \Biggr\} + O(4) \, , \label{df3} \\
\nabla ^2 \delta W ^a &=& - 4 \pi G \rho \dot{s} ^a + O(2)
  \, . \label{df4}
\end{eqnarray}
The perturbed PN Euler equation (for non-rotating, static and
spherically symmetric ground state) reads (Eq.(4.32) of Ref.
 \cite{xu01} for $\Omega = 0$)
\begin{eqnarray}\label{euler2}
0 &=& \rho ^*  \ddot{s} _a   \left( 1 + \frac{2 W}{c^2} \right)
  + \rho^* \Theta W _{,a} - \rho ^* s^b _{,a} W _{,b}
  - \rho ^*  \delta W _{,a} - \rho ^* s^b W _{,ba} \nonumber \\
 && - ( \kappa \Theta )_{,a} - ( 2 \mu s ^\beta {}_a ) _{;\beta}
  + \frac{1}{c^2} \Bigl[ - 4 \rho ^* ( \delta W _a ) _{, T}
  + \kappa \Theta W _{,a}
  + 4 ( \kappa W \Theta )_{,a} \Bigr] + O(4)\, ,
\end{eqnarray}
where $\mu$ is the shear modulus, and
\begin{eqnarray}
(2\mu s^\beta _a) _{;\beta} &=& (2\mu s _{ba}) _{,b}
  + \frac{1}{c^2} \Bigl[ - (4\mu W s_{ba} ) _{,b}
  + 4 \mu W_{,c} s_{ac} \Bigr] +O(4) \, .
\end{eqnarray}
The shear modulus $\mu $ and the compression modulus $\kappa $ are
related with the Lam\'e parameter $\lambda $ by the relation
\begin{equation}
\kappa = \lambda + 2 \mu /3  \, .
\end{equation}
The shear tensor $s_{ab}$ and the volume dilatation $\Theta$ are
given by
\begin{eqnarray}
s_{ab} &=&  \left( 1 + \frac{2W}{c^2} \right)
  \left[ s^{(a}{}_{,b)} - {1 \over 3} s^k{}_{,k} \delta _{ab}
  \right] +O(4) \, , \label{sab}\\
\Theta &=& s^k{}_{,k} + {1 \over c^2} \left( 3W_{,k} s^k
  + 3 \delta W \right)
  = \nabla\hskip-1mm\cdot\hskip-1mm {\bf s}
  + \frac{3}{c^2} (\nabla W \cdot {\bf s} + \delta W )
  +O(4) \, .
  \label{theta}
\end{eqnarray}
The gravitational scalar potential $W$ and the vector potential $W^a$
can be decomposed into a sum of two contributions (see paper
 \cite{damo91})
\begin{eqnarray}
W &=& W^+ + \overline{W} \, , \quad
W_a = W^+_a + \overline{W} _a \, , \label{decoW2}
\end{eqnarray}
where $W^+$ and $W^+_a$ are the self parts (resulting from the
gravitational action of the body under consideration) and
$\overline{W}$ and $\overline{W}_a$ are the external parts
(describing tidal and inertial forces) of the metric potentials.
Since we take the elastic mechanical ground state as an isolated
body the external potentials vanish  for this state
\begin{eqnarray}
\overline{W} &=& \overline{W}_a =0 \, , \quad
W = W^+ \, , \quad
W_a = W^+_a \, . \label{decoW5}
\end{eqnarray}
Similarly  for the perturbed state we write
\begin{eqnarray}
\delta W &=& \delta W^+ + \delta \overline{W} \,
  \quad
\delta W_a = \delta W^+_a + \delta \overline{W} _a
  \, .\label{decodW2}
\end{eqnarray}
The external parts of metric potentials, ($\delta \overline{W}$,
$\delta \overline{W}_a$), that result from the ephemerides of
external bodies are assumed to be known, so only the Eulerian
variations of the internal potentials have to be determined
self-consistently by partial differential equations. The
perturbed Euler equation can be rewritten as
\begin{eqnarray}\label{euler3}
\rho ^* f ^a
 &=& \rho ^*  \ddot{s}^a \left( 1 + \frac{4 W^+}{c^2} \right)
  + \rho^* \Theta W^+ _{,a} - \rho ^* s^b _{,a} W^+ _{,b}
  - \rho ^*  \delta W^+ _{,a} - \rho ^* s^b W^+ _{,ba} \nonumber \\
 && - ( \kappa \Theta )_{,a} -( 2 \mu s_{ba} ) _{,b}
  + \frac{1}{c^2} \Bigl[ (4 \mu W^+ s_{ba}) _{,b}
  - 4 \mu W^+_{,c} s_{ac} - 4 \rho ^* ( \delta W^+ _a ) _{, T}
   \nonumber \\
 &&  + \kappa \Theta W^+ _{,a}
  + 4 ( \kappa W^+ \Theta )_{,a} \Bigr] \, ,
\end{eqnarray}
where
\begin{equation}
f^a = \delta \overline{W}_{,a}
  + \frac{4}{c^2}(\delta \overline{W}_a)_{,T}
\end{equation}
is the external tidal force density.

To facilitate the comparison with well known results from
Newtonian theory we write the equations in a form that was
employed in Ref. \cite{wahr82}. To this end we first introduce the
post-Newtonian Cauchy elastic stress tensor ${\mathcal{T}}^{ab}$:
\begin{equation}\label{stress2}
 {\mathcal{T}}^{ab} = \kappa \Theta \delta _{ab}
  +  2 \mu s _{ab} - \frac{4W^+}{c^2} \biggl( \mu s_{ab}
  + \kappa \Theta \delta_{ab} \biggr) + O(4) \, .
\end{equation}
Then Eq.(\ref{euler3}) takes a form that can easily be compared
with the results from Wahr's paper (Ref.\cite{wahr82})
\begin{eqnarray}\label{euler5}
\left( 1 + \frac{4 W^+}{c^2} \right) \rho ^*  \ddot{\bf s}
  &=& -  \Bigl[ \rho^* \Theta \nabla W^+
  - \rho ^* \nabla W^+ \cdot (\nabla {\bf s} ) ^T
  - \rho ^*  \nabla \delta W^+
  - \rho ^* {\bf s} \cdot \nabla (\nabla W^+) \Bigr] \nonumber \\
 && + \nabla \cdot {\stackrel{\leftrightarrow}{\mathcal{T}}}
  + \frac{1}{c^2} \left\{ - \kappa \Theta  \nabla W^+
  + 2 \mu \nabla W^+ \cdot \left[ \left( \nabla {\bf s}
  + (\nabla {\bf s})^T \right) - {2 \over 3} \nabla\hskip-1mm
  \cdot\hskip-1mm{\bf s} \stackrel{\leftrightarrow}{I} \right] \right\}
  \nonumber \\
 && + \frac{4}{c^2}
  \rho ^* (\delta {\bf W}^+ ) _{,T} + \rho^* {\bf f}  + O(4) \, ,
\end{eqnarray}
where $\stackrel{\leftrightarrow}{I}$ is the second rank identity
tensor and  the superscript $T$ denotes the transpose (not to be
confused with the local time variable).

Let us now introduce  the following notation generalizing the one
from \cite{wahr82}:
\begin{equation}
\left. \begin{array}{ll}
W^+(r) = - \Phi (r) & {\rm the\; potential\; of\; the\; ground\; state}   \\
\delta W^+ = - \phi ^E _1 & {\rm the \;
  incremental\; Eulerian\; gravitational\; potential\; energy} \\
\delta \overline{W} = - \phi _T  & {\rm
  the\; tidal\; potential} \\
( \delta W^{+a})_{,T} = \stackrel{+}{A}{}^a
  & {\rm  the\; time\; derivative\; of\; the\; self\;
  part\; of\; vector\; potential} \\
(\delta {\overline{W}}^a )_{,T} \equiv
  {\overline{A}}^a & {\rm the\; time\; derivative\;
   of\; the\; external\; part\; of\; vector\; potential} \\
{\bf f} = \nabla \delta \overline{W}
  + \frac{4}{c^2}(\delta \overline{\bf W})_{,T}
 & = - \nabla \phi _T + \frac{4}{c^2} \overline{\bf A} \\
  {} & {\rm  the\; tidal\; force}
\end{array} \right\} \label{def1}
\end{equation}

\medskip\noindent
Going into Fourier space with local time variable $T$ and
replacing $\partial/\partial T$ by $i \omega$ the first perturbed
field equation (Eq.(\ref{df3})) becomes
\begin{eqnarray}
&& \left( \nabla ^2 + \frac{\omega ^2}{c^2} \right) \delta W^+
   = 4 \pi G \Bigl\{ \nabla \cdot (\rho {\bf s})
   + \frac{1}{c ^2} \Bigl[
   \nabla \cdot (p {\bf s}) + 3 \rho {\bf s} \cdot \nabla W^+
   + \rho ( \delta W^+ + \delta \overline{W} )  \nonumber \\
  && \hskip40mm + 2 \nabla \cdot (\rho W^+ {\bf s}) + 3\kappa
   \nabla \cdot {\bf s} \Bigr] \Bigr\} + O(4) \, , \label{df5}
\end{eqnarray}
since the tidal potential $\delta \overline{W}$ satisfies
D'Alembert's equation. \\
The second perturbed field equation (Eq.(\ref{df4})), for the same
reason, takes the form
\begin{equation}\label{df7}
 \nabla ^2 ( \delta W^{+a} )_{,T} \equiv
  \nabla ^2 \stackrel{+}{A}{}^a
  = 4 \pi G \omega^2 \rho s^a + O(2) \, .
\end{equation}
By using the definition above (Eqs.(\ref{def1})), the
Eq.(\ref{df5}) can be rewritten as
\begin{eqnarray}
- \left( \nabla ^2 + \frac{\omega ^2}{c^2} \right) \phi^E_1
   &=& 4 \pi G \Bigl\{ \nabla \cdot (\rho^* {\bf s})
   + \frac{1}{c ^2} \Bigl[
   - 5 \rho^* {\bf s} \cdot \nabla \Phi
   - \rho^* \phi ^E_1 - \rho^* \phi_T  \nonumber \\
  &&  - 2 \Phi \nabla \cdot (\rho^* {\bf s})
  + 3 \kappa \nabla \cdot {\bf s}
   \Bigr] \Bigr\} + O(4) \, . \label{df9}
\end{eqnarray}
The perturbed Euler equation (Eq.(\ref{euler5})) takes the form
\begin{eqnarray}\label{euler6}
&& - \left( 1 - \frac{4 \Phi}{c^2} \right)
  \rho ^* \omega ^2 {\bf s}
   = - \Bigl[ \rho ^*  \nabla \phi ^E _1
  + \rho ^* \nabla \Phi \cdot ( \nabla {\bf s} ) ^T
  + \rho ^*  {\bf s} \cdot \nabla ( \nabla \Phi ) \nonumber \\
 && \hskip16mm - \rho ^* \Theta \nabla \Phi \Bigr]
  + \nabla \cdot {\stackrel{\leftrightarrow}{\mathcal{T}}}
  + \frac{1}{c^2} \left\{ \kappa \Theta  \nabla \Phi
  - 2 \mu \nabla \Phi \cdot \left[ \left( \nabla {\bf s}
  + (\nabla {\bf s})^T \right) - {2 \over 3} \nabla
  \cdot {\bf s} \stackrel{\leftrightarrow}{I} \right] \right\}
  \nonumber \\
 && \hskip16mm + \frac{4}{c^2}
  \rho ^* ( \stackrel{+}{\bf A} + \overline{\bf A} )
  - \rho^* \nabla \phi _T  + O(4) \, ,
\end{eqnarray}
where
\begin{equation}\label{stress4}
{\mathcal{T}}^{ab}
  =  \kappa \Theta \delta _{ab} +  2 \mu s _{ab}
  + \frac{4\Phi}{c^2} \Bigl( \mu s_{ab}
  + \kappa \Theta \delta _{ab} \Bigr) + O(4) \, ,
\end{equation}
and
\begin{equation}\label{theta1}
\Theta = \nabla\hskip-1mm\cdot\hskip-1mm {\bf s}
  - \frac{3}{c^2}\left( \nabla \Phi \cdot {\bf s}
  + \phi^E_1 + \phi _T \right) + O(4)  \, .
\end{equation}
Eqs.(\ref{df7}), (\ref{df9}), (\ref{euler6}) and (\ref{stress4})
are the ones that  will now be expanded in
terms of generalized spherical harmonics (scalar-, vector- and
tensor spherical harmonics).


\section{Generalized spherical harmonics}

Eqs.(\ref{df7})---(\ref{stress4}) are complicated partial
differential equations (PDE) that will be turned into a set of
ordinary differential equations (ODE) by means of expansions in
terms of scalar-, vector- and tensor spherical harmonics or,
briefly, in terms of generalized spherical harmonics that are
described e.g., by Phinney \& Burridge
 \cite{phin73}, Smith \cite{smit74} and Wahr \cite{wahr82}.
 Here tensors of arbitrary rank are represented by their
components along the complex basis ${\bf e} _-$, ${\bf e} _0$,
${\bf e}_+$, where
\begin{equation}
{\bf e}_- = \frac{1}{\sqrt{2}} ({\bf e}_ \theta - i {\bf e}_\phi )
 \, , \quad
{\bf e}_0 = {\bf e}_r \, , \quad {\bf e}_+ = - \frac{1}{\sqrt{2}}
( {\bf e}_ \theta + i {\bf e}_\phi )
\end{equation}
and ${\bf e} _r$, ${\bf e} _\theta$ and ${\bf e}_\phi$ are (Euclidean) unit
vectors in $r$, $\theta$ and $\phi$ direction. Components of
a tensor field along the complex basis are called canonical
components and the basis itself is called canonical basis.
Generalized spherical harmonics (GSH) are denoted by $D^l_{mn}
(\theta, \phi)$ and defined by
\begin{equation}
D^l _{mn} (\theta, \phi) = (-1) ^{m+n} P_l ^{nm} (\cos \theta)
  \exp (im \phi) \, ,
\end{equation}
where $P_l ^{nm}$ is a generalized associated Legendre function
\begin{eqnarray}
P_l^{nm} (x) &=& \frac{(-1)^{l-n}}{2^l (l-n) !} \left[
  \frac{(l-n)! (l+m)!}{(l+n)! (l-m)!} \right] ^{\frac{1}{2}}
  (1-x)^{(n-m)/2} (1+x) ^{-(m+n)/2} \nonumber \\
  && \times \left( \frac{d}{dx} \right) ^{l-m} \left[
  (1-x)^{l-n} (1+x)^{l+n} \right] \, .
\end{eqnarray}
$D^l _{mn}$ are defined only for $l \geq 0$ and $| m | \leq l $
(see, e.g., \cite{phin73} and \cite{smit74} for a complete
discussion).

By using GSH, a scalar, vector and 2nd rank tensor can be expanded
as
\begin{eqnarray}
\phi (r, \theta, \phi ) &=& \sum_{l=0} ^{\infty} \sum_{m=-l} ^l
  \phi^m_l (r) D^l _{m0} (\theta , \phi) \, ,\label{expS} \\
{\bf u}(r, \theta, \phi ) &=& \sum_{n =-1} ^{+1}
  \sum_{l=0}^{\infty} \sum_{m=-l} ^l u^{mn}_l (r)
  D^l _{mn} (\theta, \phi) {\bf e}_n \, ,\label{expV} \\
\stackrel{\leftrightarrow}{\mathcal{T}} (r, \theta, \phi )
  &=& \sum_{a , b = -1} ^{+1} \sum_{l=0} ^{\infty}
  \sum_{m=-l} ^l T^{m a b}_l (r)
  D^l _{m(a + b)}(\theta , \phi) {\bf e}_a
  {\bf e}_b  \, .\label{expT}
\end{eqnarray}
The $D^l _ {m0}$ are proportional to the usual spherical harmonics
($D^l _ {m0} = (-1) ^m \sqrt{ 4 \pi /(2l+1)} Y ^m_l$) and the  $D^l
_{mn} (\theta, \phi)$ are similar to the $Y^{mn}_l (\theta ,\phi)$
of paper \cite{phin73} ($D^l _{mn} = (-1)^{m+n} Y_l^{mn} $).
Some useful formulas for objects involving GSH can be found in the
Appendix.


\section{Post-Newtonian perturbation equations}

In this section we will rewrite Eqs.(\ref{df7}), (\ref{df9}),
(\ref{euler6}) and (\ref{stress4}) in a compact form by using GSH.
First we expand the scalars, vectors and tensors appearing in these
equations in the following way
\begin{eqnarray}
\phi_1 ^E (r, \theta, \phi ) &=& \sum_{l=0} ^{\infty}
 \sum_{m=-l}^l \phi^m_l (r) D^l _{m0}
  (\theta , \phi) \, ,\label{gsh1}\\
\phi_T (r, \theta, \phi ) &=& \sum_{l=2} ^{\infty} \sum_{m=-l} ^l
  {}^T\hskip-1mm\phi^m_l (r) D^l _{m0} (\theta , \phi)\label{gsh2} \\
s^a (r, \theta, \phi ) &=&  \sum_{l=0} ^{\infty} \sum_{m=-l} ^l
  s^{m a}_l (r) D^l _{m a} (\theta , \phi)  \, ,
  \label{gsh3} \\
\stackrel{+}{A}{}^a (r, \theta, \phi ) &=&
  \sum_{l=0}^{\infty} \sum_{m=-l} ^l
  \stackrel{+}{A}{}^{m a}_l (r)
  D^l _{m a} (\theta , \phi)  \, ,\label{gsh4} \\
\overline{A}^a (r, \theta, \phi ) &=&
  \sum_{l=0}^{\infty} \sum_{m=-l} ^l
  \overline{A}^{m a}_l (r)
  D^l _{m a} (\theta , \phi)  \, ,\label{gsh5} \\
\mathcal{T}^{ab} (r, \theta, \phi )
 &=& \sum_{l=0} ^{\infty} \sum_{m=-l} ^l
  T^{m a b}_l (r) D^l _{m(a + b)}
  (\theta , \phi)  \, . \label{gsh6}
\end{eqnarray}
Here $s^a$ denotes the displacement field, $\mathcal{T}^{ab}$ the
incremental Cauchy elastic stress tensor ($a , \, b = -1 , \, 0,
\, +1 $). The gravitational potential $\Phi$ of the ground
(equilibrium) state should not be expanded, since it is a known
function of the radial coordinate $r$.


\subsection{Field equations}

By using the expansions from
Eqs.(\ref{gsh1})---(\ref{gsh3}), the formulas given in the
Appendix and the orthogonality properties of $D^l _{mn}$ for each
appropriate value of  $l,~m$ and $n$, the Eq.(\ref{df9}) reduces
to
\begin{eqnarray}
&&  \left( \frac{d^2}{dr^2}
  + \frac{2}{r}\frac{d}{dr} - \frac{l(l+1)}{r^2}
  + \frac{\omega ^2}{c^2} \right) \phi^m _l \nonumber \\
&& \hskip6mm = -4 \pi G
  \left[ \left( \frac{d}{dr} + \frac{2}{r}\right) (\rho^*
  s_l^{m0})
  +\frac{\rho^*}{r} \sqrt{\frac{l(l+1)}{2}} \left(
  s_l^{m+} + s_l^{m-} \right) \right]  \nonumber \\
&& \hskip10mm
  + \frac{4\pi G}{c^2} \left[ 5\rho^* (\frac{d}{dr} \Phi) s_l^{m0}
  + \rho^* (\phi ^m_l + {}^T \phi ^m_l )
  + (2\rho^* \Phi - 3\kappa ) \left(
  \frac{d}{dr} + \frac{2}{r} \right) s_l^{m0} \right. \nonumber \\
&& \hskip10mm  \left. +(2\rho^* \Phi - 3\kappa )
  \frac{1}{r} \sqrt{\frac{l(l+1)}{2}}
  (s_l^{m+} + s_l^{m-} ) + 2 \Phi (\frac{d}{dr} \rho^* ) s_l^{m0}
  \right] \, .  \label{df10}
\end{eqnarray}

To write this equation in a more compact form it is useful to define
new scalar functions (\cite{wahr82})
\begin{equation}\label{def2}
\left. \begin{array}{l}
 U^m_l = s^{m0}_l \\
 V^m_l = s^{m+}_l + s^{m-}_l  \\
 W^m_l = s^{m+}_l - s^{m-}_l  \\
 P^m_l = T^{m00}_l  \\
 Q^m_l = T^{m0+}_l + T^{m0-}_l  \\
 R^m_l = T^{m0+}_l - T^{m0-}_l
 \end{array} \right\} \, .
\end{equation}
Let
\begin{equation}\label{eq1}
g^m_l = \frac{d}{dr}  \phi^m_l + 4\pi G\rho^* U^m_l \, ,
\end{equation}
then
\begin{equation}\label{eq1a}
\frac{d^2}{dr^2} \phi^m_l = \frac{d}{dr} g^m_l
  - 4\pi G \frac{d}{dr} (\rho^* U^m_l)  \, .
\end{equation}
Substituting Eqs.(\ref{def2}), (\ref{eq1}) and (\ref{eq1a}) into
Eq.(\ref{df10}), we get
\begin{eqnarray}
\frac{d}{dr} g^m_l &=& \left( \frac{l(l+1)}{r^2}
  - \frac{\omega ^2}{c^2} \right) \phi^m _l
  -\frac{2}{r} g^m_l - \frac{4\pi G}{r} \rho^* L_1 V^m_l
  + \frac{4\pi G}{c^2} \Biggl\{ 5\rho^* g_0 U_l^m
  + \rho^* ( \phi ^m_l + {}^T \phi ^m_l) \nonumber \\
  && + (2\rho^* \Phi - 3\kappa ) \left[ \left(
  \frac{d}{dr} + \frac{2}{r} \right) U_l^m
  + \frac{L_1}{r} V_l^m \right] + 2\Phi \left( \frac{d}{dr} \rho^*
  \right) U_l^m \Biggr\} \, ,  \label{eq2}
\end{eqnarray}
where
\begin{eqnarray}
g_0 &\equiv & \frac{d}{dr} \Phi \, , \label{g0}\\
L_1 & \equiv & \sqrt{\frac{l(l+1)}{2}} \, .\label{L_1}
\end{eqnarray}
In the Newtonian limit Eq.(\ref{eq1}) and Eq.(\ref{eq2}) are
correspond to Eq.(III.33) and Eq.(III.34) of \cite{wahr82}
respectively.

\bigskip
Next  we consider the second perturbed field equation
Eq.(\ref{df7}). Expand $\stackrel{+}{A}{}^{(-/0/+)}$ according to
Eq.(\ref{gsh4}), i.e.,
\begin{equation}
\stackrel{+}{A}{}^{(-/0/+)} (r, \theta, \phi ) =  \sum_{l=0}
^{\infty} \sum_{m=-l} ^l
  \stackrel{+}{A}{}^{m (-/0/+)}_l (r) D^l _{m -1} (\theta , \phi) \, .
\end{equation}
For the explicit calculation we have
\begin{equation} \label{naA+}
\left( \nabla ^2 \stackrel{+}{\bf A}{} \right) ^a = \left( \nabla
\cdot \nabla \stackrel{+}{\bf A} \right) ^a
   \equiv \xi ^a \, .
\end{equation}
Since $\nabla \stackrel{+}{\bf A}$ is not a symmetric second rank
tensor so that  we cannot use Phinney's formulas directly. We
obtain (for more details see the Appendix)
\begin{eqnarray}
\xi ^0 &=& \sum_{l=0}^{\infty} \sum_{m=-l} ^l \left\{ \left[
  \frac{d^2}{dr^2}+ \frac{2}{r} \frac{d}{dr}
  - \frac{2}{r^2} (1+ L_1^2)\right] \stackrel{+}{A}{}^{m0}_l
  - \frac{2 L_1}{r^2} (\stackrel{+}{A}{}^{m+}_l +
  \stackrel{+}{A}{}^{m-}_l ) \right\} D^l_{m0}
  \, , \label{xi01}  \\
\xi ^- &=& \sum_{l=0}^{\infty} \sum_{m=-l} ^l
  \left\{ \left[ \frac{d^2}{dr^2} + \frac{2}{r} \frac{d}{dr}
  - \frac{2}{r^2} (1 + L_2^2)\right] \stackrel{+}{A}{}^{m-}_l
  - \frac{2 L_1}{r^2} \stackrel{+}{A}{}^{m0}_l  \right\} D^l_{m-1}
  \, , \label{xi-1}  \\
\xi ^+ &=& \sum_{l=0}^{\infty} \sum_{m=-l} ^l
  \left\{ \left[ \frac{d^2}{dr^2} + \frac{2}{r} \frac{d}{dr}
  - \frac{2}{r^2} (1 + L_2^2)\right] \stackrel{+}{A}{}^{m+}_l
  - \frac{2L_1}{r^2} \stackrel{+}{A}{}^{m0}_l  \right\} D^l_{m+1}
  \, ,  \label{xi+1}
\end{eqnarray}
where $L_1$ is defined in Eq.(\ref{L_1}), and
\begin{equation}\label{L_2}
L_2  \equiv  \sqrt{\frac{(l-1)(l+2)}{2}} \, .
\end{equation}
The expansion  of $s^a$ takes the form
\begin{equation}
s^{(-/0/+)} (r, \theta, \phi ) =  \sum_{l=0} ^{\infty}
  \sum_{m=-l}^l
  s^{m (-/0/+)}_l (r) D^l _{m (-1/0/+1)} (\theta , \phi)\, .
\end{equation}
Using this expansion in the field equations (\ref{df7}) and the
orthogonality relations of $D^l_{mn}$ one finds
\begin{eqnarray}
&& \left[ \frac{d^2}{dr^2} + \frac{2}{r} \frac{d}{dr}
  - \frac{2}{r^2} (1 + L_1^2)\right] \stackrel{+}{B}{}^m_l
  - \frac{2 L_1}{r^2} \stackrel{+}{E}{}^m_l
  = 4\pi G \omega ^2 \rho U^m_l \, , \label{pf6} \\
&& \left[ \frac{d^2}{dr^2} + \frac{2}{r} \frac{d}{dr}
  - \frac{2}{r^2} (1+ L_2^2)\right] \stackrel{+}{E}{}^m_l
  - \frac{4 L_1}{r^2} \stackrel{+}{B}{}^m_l
  = 4 \pi G \omega^2\rho V^m_l  \, , \label{pf7}  \\
&& \left[ \frac{d^2}{dr^2} + \frac{2}{r} \frac{d}{dr}
  - \frac{2}{r^2} (1 + L_2^2)\right] \stackrel{+}{H}{}^m_l
  = 4 \pi G \omega^2\rho W^m_l  \, . \label{pf8}
\end{eqnarray}
with the definition
\begin{equation}\label{def3}
\left. \begin{array}{l}
\stackrel{+}{A}{}^{m0}_l = \stackrel{+}{B}^m_l  \\
\stackrel{+}{A}{}^{m+}_l + \stackrel{+}{A}{}^{m-}_l
  = \stackrel{+}{E}^m_l  \\
\stackrel{+}{A}{}^{m+}_l - \stackrel{+}{A}{}^{m-}_l
  = \stackrel{+}{H}^m_l  \\
\overline{A}^{m0}_l = \overline{B}^m_l  \\
\overline{A}^{m+}_l + \overline{A}^{m-}_l
  = \overline{E}^m_l  \\
\overline{A}^{m+}_l - \overline{A}^{m-}_l
  = \overline{H}^m_l
 \end{array} \right\}
\end{equation}
(The definitions for $\overline{B}^m_l $, $\overline{E}^m_l $
and $\overline{H}^m_l $ will be used in Section IV.C.)
For numerical applications it is useful to reduce the set of
seconder order differential equations
(Eq.(\ref{pf6})---(\ref{pf8})) to an equivalent set of first order.
With
\begin{eqnarray}
\frac{d}{dr} \stackrel{+}{B}{}^m_l &=& b^m_l \, , \label{eq3} \\
\frac{d}{dr} \stackrel{+}{E}{}^m_l &=& e^m_l \, , \label{eq4} \\
\frac{d}{dr} \stackrel{+}{H}{}^m_l &=& h^m_l \, , \label{eq5}
\end{eqnarray}
the post-Newtonian field equations, Eqs.(\ref{pf6}) - (\ref{pf8}),
finally take the form
\begin{eqnarray}
&& \frac{d}{dr} b^m_l = - \frac{2}{r} b^m_l
  + \frac{2}{r^2}(1+ L_1^2) \stackrel{+}{B}{}^m_l
  + \frac{2L_1}{r^2} \stackrel{+}{E}{}^m_l
  + 4 \pi G \omega ^2 \rho U^m_l \, , \label{eq6} \\
&& \frac{d}{dr} e^m_l = - \frac{2}{r} e^m_l
  + \frac{2}{r^2}(1 + L_2^2) \stackrel{+}{E}{}^m_l
  + \frac{4L_1}{r^2} \stackrel{+}{B}{}^m_l
  + 4 \pi G \omega ^2 \rho V^m_l \, , \label{eq7} \\
&& \frac{d}{dr} h^m_l = - \frac{2}{r} h^m_l
  + \frac{2}{r^2}(1 + L_2^2) \stackrel{+}{H}{}^m_l
  + 4 \pi G \omega ^2 \rho W^m_l \, . \label{eq8}
\end{eqnarray}
Compared with the Newtonian case we have six additional  equations
(Eqs.(\ref{eq3}---\ref{eq8})) and six additional unknown functions
$b_l^m$, $e_l^m$, $h_l^m$, $\stackrel{+}{B}{}_l^m$,
$\stackrel{+}{E}{}_l^m$, $\stackrel{+}{H}{}_l^m$ that only appear
in the post-Newtonian formalism.


\subsection{Stress-strain relation}

According to  Eqs.(\ref{stress4}), (\ref{theta1}) and (\ref{sab}),
and the stress tensor takes the form
\begin{eqnarray}
\stackrel{\leftrightarrow}{\mathcal{T}} &=& \lambda
\nabla\hskip-1mm\cdot\hskip-1mm {\bf s}
  \stackrel{\leftrightarrow}{I}
  + \mu \left[ \nabla {\bf s} + (\nabla {\bf s})^T \right]
  + \frac{\kappa }{c^2} \left[ 4 \Phi
  \nabla\hskip-1mm\cdot\hskip-1mm {\bf s}
  -3 (\nabla \Phi \cdot {\bf s} + \phi^E_1 +\phi_T ) \right]
  \stackrel{\leftrightarrow}{I} \, . \label{stress6}
\end{eqnarray}
Using the GSH expansion for the Cauchy elastic stress-tensor one
finds that
\begin{eqnarray}
\mathcal{T}{}^{ab} (r, \theta, \phi )
  &=& \sum_{l=0} ^{\infty}
  \sum_{m=-l} ^l T^{m a b}_l (r)
  D^l _{m(a + b)}(\theta , \phi) \nonumber \\
 &=& \lambda \nabla\hskip-1mm\cdot\hskip-1mm {\bf s} e^{ab}
  + \mu ( s^{a,b} + s^{b,a} )
  + \frac{\kappa }{c^2}  \left[ 4 \Phi
  \nabla\hskip-1mm\cdot\hskip-1mm {\bf s}
  -3 (\nabla \Phi \cdot {\bf s}
  + \phi^E_1 + \phi_T ) \right]  e^{ab}
   \, , \label{stress7}
\end{eqnarray}
where $e_{ab}$ and $e^{ab}$ are the canonical components of the
identity tensor and are equal to
\begin{equation}\label{eab}
( e_{ab} )
  = ( e^{ab} )
  = \left( \begin{array}{ccc}
 0 & 0 & -1 \\
 0 & 1 & 0 \\
 -1 & 0 & 0
 \end{array} \right)
 \hskip1cm a , \; b = - , \; 0 , \;  +
\end{equation}
Using
\begin{eqnarray}
&& \nabla\hskip-1mm\cdot\hskip-1mm {\bf s} = s^{c,d}e_{cd}
  = \sum_{l=0} ^{\infty} \sum_{m=-l} ^l
  \left[ \left( \frac{d}{dr} + \frac{2}{r} \right) s^{m0}_l
  + \frac{1}{r} \sqrt{\frac{l(l+1)}{2}}
  ( s^{m+}_l + s^{m-}_l ) \right] D^{m0}_l \, , \label{divs} \\
&& \nabla \Phi \cdot {\bf s} = \Phi ^{,c} s^d e_{cd}
  = \Phi ^{,0} s^0 \, , \label{sdiv}
\end{eqnarray}
and the orthogonality properties of $D^l_{mn}$ we get
\begin{eqnarray}
T^{m--}_l &=& - \frac{2\mu}{r} L_2 s^{m-}_l  \, ,
  \label{t--} \\
T^{m0-}_l &=& T^{m-0}_l = \mu \left[ \left(
  \frac{d}{dr} - \frac{1}{r} \right)
  s^{m-}_l - \frac{1}{r} L_1 s^{m0}_l  \right]  \, ,
  \label{t0-} \\
T^{m00}_l &=& \left( \lambda + \frac{4\kappa}{c^2} \Phi \right)
  \left[ \left( \frac{d}{dr} + \frac{2}{r} \right) s^{m0}_l
  + \frac{L_1}{r}
  ( s^{m+}_l + s^{m-}_l ) \right]  \nonumber \\
 && + 2\mu \left( \frac{d}{dr} s^{m0}_l \right)
  - \frac{3\kappa}{c^2}
  \left( ( \frac{d}{dr} \Phi ) s^{m0}_l + \phi ^m_l
  + {}^T \phi^m_l \right) \, .  \label{t00} \\
T^{m+-}_l &=& T^{m-+}_l = - \left( \lambda
  + \frac{4\kappa }{c^2} \Phi \right) \left[ \left(
  \frac{d}{dr} + \frac{2}{r} \right) s^{m0}_l
  + \frac{L_1}{r} ( s^{m+}_l + s^{m-}_l ) \right]  \nonumber \\
 && - \mu \left[ \frac{L_1}{r}
  ( s^{m+}_l + s^{m-}_l ) + \frac{2}{r} s^{m0}_l \right]
  + \frac{3\kappa}{c^2}
  \left( ( \frac{d}{dr} \Phi ) s^{m0}_l + \phi ^m_l
  + {}^T \phi^m_l  \right) \, .  \label{t+-} \\
T^{m0+}_l &=& T^{m+0}_l = \mu \left[ \left( \frac{d}{dr} -
\frac{1}{r} \right)
  s^{m+}_l - \frac{L_1}{r} s^{m0}_l \right]  \, ,
  \label{t0+} \\
T^{m++}_l &=& - \frac{2\mu}{r} L_2 s^{m+}_l \, .\label{t++}
\end{eqnarray}
The relation for $T_l^{m00} \equiv P^m_l$ can be written in the
form
\begin{eqnarray}
&& \left[ \left( \lambda + 2\mu \right)
  + \frac{4\kappa }{c^2} \Phi \right] \frac{d}{dr} U^m_l
  = P^m_l - \left( \lambda + \frac{4\kappa }{c^2} \Phi \right)
  \left( \frac{2}{r} U^m_l + \frac{L_1}{r} V^m_l \right)
  + \frac{3\kappa}{c^2}( g_0 U^m_l+ \phi^m_l + {}^T \phi ^m_l )
   \, ,\nonumber \\
 &&  \label{ulm}
\end{eqnarray}
which is the post-Newtonian version of Eq.(III.35) from Wahr
(1982).

By means of Eqs.(\ref{t0-}) and (\ref{t0+}) we get
\begin{eqnarray}
\mu \frac{d}{dr} V^m_l &=&
  Q^m_l + \frac{\mu}{r} ( V^m_l + 2 L_1 U^m_l ) \, ,
  \label{vlm} \\
\mu \frac{d}{dr} W^m_l &=& R^m_l + \frac{\mu}{r} W^m_l
  \, ,\label{wlm}
\end{eqnarray}
that generalizes Wahr's Eqs.(III.36) and (III.37) to the
post-Newtonian level. Eqs.(\ref{ulm}), (\ref{vlm}) and (\ref{wlm})
are the post-Newtonian version of the stress-strain relation.


\subsection{Euler equation (Conservation of momentum)}

According to Eqs.(\ref{euler6}) and (\ref{theta1}), in the complex
basis $ {\bf e}_-,~{\bf e}_0,~{\bf e}_+$ the perturbed Euler
equation can be expressed as
\begin{eqnarray}
&& - \left( 1 - \frac{4 \Phi}{c^2} \right)
  \rho ^* \omega ^2 s^a =
  - \Biggl\{ \rho ^*  \phi ^{E,a}_1
  + \rho ^*  \Phi^{,b} s^{c,a} e_{bc}
  + \rho ^* s^b \Phi^{,ca} e_{bc} \nonumber \\
 && \hskip16mm - \rho ^* \left[
  (\nabla\hskip-1mm\cdot\hskip-1mm {\bf s})
  - \frac{3}{c^2} \left( \Phi^{,c} s^d e_{cd}
  + \phi^E_1 +\phi_T \right) \right] \Phi ^{,a} \Biggr\}
  + \zeta ^a - \rho^* \phi ^{,a}_T
  + \frac{4}{c^2}\rho^* ( \stackrel{+}{A}{}^a
  + \overline{A} ^a ) \nonumber \\
 && \hskip16mm + \frac{1}{c^2} \Biggl\{ \kappa\Phi^{,a}
  (\nabla\hskip-1mm\cdot\hskip-1mm {\bf s})
  - 2 \mu \left[ \Phi^{,g} \left( s^{a,b}
  + s^{b,a} \right) e_{g b}
  - {2 \over 3} \Phi^{,a} (\nabla\hskip-1mm\cdot\hskip-1mm {\bf s} )
  \right] \Biggr\} \, , \label{euler8}
\end{eqnarray}
where $\zeta ^a = \left( \nabla \cdot
\stackrel{\leftrightarrow}{\mathcal{T}} \right) ^a =
{\mathcal{T}}^{ab,c} e_{bc}$. This relation comprises the three
equations for $a = (-,0,+)$.

\bigskip
By means of Eqs.(\ref{t--}) and (\ref{t+-}), the $a = - $ equation
can be written in the form
\bigskip
\begin{eqnarray}
\frac{d}{dr} T^{m0-}_l &=& - \frac{3}{r} T^{m0-}_l
  + \frac{2\mu}{r^2} L_2^2 s^{m-}_l
  + \frac{L_1}{r} \left\{ \frac{\mu+\lambda}{r} \left[
  L_1 (s^{m+}_l + s^{m-}_l ) + 2 s^{m0}_l \right]
  + \lambda \frac{d}{dr} s^{m0}_l \right\} \nonumber \\
 && -\frac{L_1}{r}\rho^* ( g_0 s^{m0}_l + \phi^m_l
  + {}^T\hskip-1mm\phi^m_l )
  - \rho^* \omega^2 s^{m-}_l \nonumber \\
 && + \frac{1}{c^2} \Biggl\{
  \frac{4\kappa}{r} \Phi L_1 \left[ \frac{L_1}{r}
  (s^{m+}_l + s^{m-}_l ) + \left( \frac{d}{dr}
  + \frac{2}{r} \right) s^{m0}_l \right] \nonumber \\
 &&  - \frac{3\kappa}{r} L_1 \left(
  g_0 s^{m0}_l + \phi^m_l + {}^T \phi^m_l \right)
  + 4 \Phi \rho^* \omega^2 s^{m-}_l  \nonumber \\
 && + 2 \mu g_0 \left[ \left( \frac{d}{dr}-\frac{1}{r}
  \right) s^{m-}_l - \frac{1}{r} L_1 s^{m0}_l \right]
  - 4 \rho^* ( \stackrel{+}{A}{}^{m-}_l + \overline{A}^{m-}_l )
  \Biggr\} \, . \label{euler-}
\end{eqnarray}

\bigskip
The $a = 0$ equation, by using Eq.(\ref{t+-}), can be written as
\begin{eqnarray}
&& \frac{d}{dr} T^{m00}_l = - \frac{2}{r} T^{m00}_l
  - \frac{L_1}{r} (T^{m0+}_l +T^{m0-}_l )
  + \frac{2}{r} \left\{ \frac{\mu+\lambda}{r} \left[
  L_1 (s^{m+}_l + s^{m-}_l ) + 2 s^{m0}_l \right]
  + \lambda \frac{d}{dr} s^{m0}_l \right\} \nonumber \\
&& \hskip10mm - \rho^* \omega^2 s^{m0}_l + \rho^* \left(
  \frac{d}{dr} \phi^m_l + 4\pi G \rho^* s^{m0}_l \right)
  - \frac{4}{r}\rho^* g_0 s^{m0}_l
  -\frac{L_1}{r}\rho^* g_0 (s^{m+}_l + s^{m-}_l )
  + \rho \frac{d}{dr} \left( {}^T\hskip-1mm\phi^m_l \right)
   \nonumber \\
&& \hskip10mm + \frac{1}{c^2} \Biggl\{ \left[
  \frac{8\kappa}{r} \Phi
   - ( \lambda + 2\mu ) g_0 \right] \left(
   \frac{d}{dr} + \frac{2}{r} \right) s^{m0}_l
  + \left[ \frac{8\kappa}{r} \Phi
  - (\lambda + 2\mu ) g_0 \right]
  \frac{L_1}{r} ( s^{m+}_l + s^{m-}_l ) \nonumber \\
&& \hskip10mm  + 4\mu g_0 \frac{d}{dr} s^{m0}_l
  - 4\rho^* ( \stackrel{+}{A}{}^{m0}_l
  + \overline{A}^{m0}_l )
  + 3 \left( \rho^* g_0 - \frac{2\kappa}{r} \right)
  ( g_0 s^{m0}_l + \phi ^m_l +{}^T \phi^m_l ) \nonumber \\
&& \hskip10mm + 4\Phi \rho^* \omega^2 s^{m0}_l
  + 8 \pi G ( p - \rho^* \Phi ) \Biggr\}  \, , \label{euler0}
\end{eqnarray}
where we have used Eq.(\ref{field3}). The pressure $p$ is
determined from the equation of state $p=p(\rho )$ that is assumed
to be a known function inside the reference body.

\bigskip
Finally,  the $a = +$ equation can be brought into the form
\begin{eqnarray}
\frac{d}{dr} T^{m0+}_l &=& - \frac{3}{r} T^{m0+}_l
  + \frac{2\mu}{r^2} L_2^2 s^{m+}_l
  + \frac{L_1}{r} \left\{ \frac{\mu+\lambda}{r} \left[
  L_1 (s^{m+}_l + s^{m-}_l ) + 2 s^{m0}_l \right]
  + \lambda \frac{d}{dr} s^{m0}_l \right\} \nonumber \\
 && -\frac{L_1}{r}\rho^* ( g_0 s^{m0}_l + \phi^m_l
  + {}^T\hskip-1mm\phi^m_l )
  - \rho^* \omega^2 s^{m+}_l \nonumber \\
 && + \frac{1}{c^2} \Biggl\{
  \frac{4\kappa }{r} \Phi L_1
 \left[ \frac{L_1}{r} (s^{m+}_l + s^{m-}_l )
  + \left( \frac{d}{dr} + \frac{2}{r} \right) s^{m0}_l \right]
  \nonumber \\
 &&  - \frac{3\kappa }{r} L_1  (g_0 s^{m0}_l + \phi^m_l
  + {}^T \phi^m_l )
  + 4 \Phi \rho^* \omega^2 s^{m+}_l  \nonumber \\
 && + 2 \mu g_0 \left[ \left( \frac{d}{dr}-\frac{1}{r}
  \right) s^{m+}_l - \frac{1}{r} L_1 s^{m0}_l \right]
  - 4 \rho^* ( \stackrel{+}{A}{}^{m+}_l + \overline{A}^{m+}_l )
  \Biggr\} \, . \label{euler+}
\end{eqnarray}

\bigskip
We will also rewrite these PN Euler equations to facilitate the
comparison with the corresponding Newtonian equations. With the
definitions from Eqs.(\ref{def2}), (\ref{eq1}) and (\ref{def3}),
the $a = 0$ Equation (\ref{euler0}) can be rewritten as
\bigskip
\begin{eqnarray}
\frac{d}{dr} P^m_l &=& - \frac{2}{r} P^m_l
  - \frac{L_1}{r} Q^m_l
  + \frac{2}{r} \left[ \frac{\mu+\lambda}{r} \left(
  L_1 V^m_l + 2 U^m_l \right)
  + \lambda \frac{d}{dr} U^m_l \right] \nonumber \\
  && - \rho^* \omega^2 U^m_l + \rho^* g^m_l
  - \frac{4}{r}\rho^* g_0 U^m_l
  -\frac{L_1}{r}\rho^* g_0 V^m_l
  + \rho^* \frac{d}{dr} ( {}^T\hskip-1mm\phi^m_l )  \nonumber \\
  && + \frac{1}{c^2} \Biggl\{
   \left[ \frac{8\kappa }{r} \Phi -(\lambda + 2\mu) g_0
   \right] \left[ \left( \frac{d}{dr} + \frac{2}{r} \right) U^m_l
  + \frac{L_1}{r} V^m_l \right] + 4\mu g_0 \frac{d}{dr}U^m_l
   \nonumber \\
  && + 3 \left( \rho^* g_0 - \frac{2\kappa}{r} \right)
   ( g_0 U^m_l + \phi^m_l + {}^T \phi^m_l)
  + 4\Phi \rho^* \omega^2 U^m_l + 8 \pi G (p - \rho^* \Phi )
  \nonumber \\
  && - 4\rho^* ( \stackrel{+}{B}{}^m_l + \overline{B}^m_l )
  \Biggr\} \, , \label{plm}
\end{eqnarray}
generalizing Wahr's Equation (III.30).

Similarly, combining Eq.(\ref{euler+}) with (\ref{euler-}) we get
\medskip
\begin{eqnarray}
\frac{d}{dr} Q^m_l &=& - \frac{3}{r} Q^m_l
  + \frac{2\mu}{r^2} L_2^2 V^m_l + \frac{2L_1}{r} \left[
  \frac{\mu+\lambda}{r} \left( L_1 V^m_l + 2 U^m_l \right)
  + \lambda \frac{d}{dr} U^m_l \right] \nonumber \\
  && - \rho^* \omega^2 V^m_l
  -\frac{2L_1}{r}\rho^* ( g_0 U^m_l +\phi^m_l
  + {}^T\hskip-1mm\phi^m_l ) \nonumber \\
  && + \frac{1}{c^2} \Biggl\{
  \frac{8\kappa }{r} \Phi L_1
  \left[ \frac{L_1}{r} V^m_l + \left( \frac{d}{dr}
  + \frac{2}{r} \right) U^m_l \right]
  - \frac{6\kappa }{r} L_1 (g_0 U^m_l
  + \phi^m_l + {}^T \phi^m_l ) \nonumber \\
  && + 2 \mu g_0 \left[ \left( \frac{d}{dr}
  - \frac{1}{r} \right) V^m_l
  - \frac{2L_1}{r} U^m_l \right]
  + 4 \Phi \rho^* \omega^2 V^m_l - 4 \rho^*
  ( \stackrel{+}{E}{^m_l} + \overline{E}^m_l ) \Biggr\}
  \, , \label{qlm}
\end{eqnarray}
which is the post-Newtonian version of Wahr's Eq.(III.31).

\bigskip
Wahr's
Eq.(III.32) is generalized if we combine Eq.(\ref{euler+})
with (\ref{euler-}) (using Eq.(\ref{wlm}))
\begin{eqnarray}
\frac{d}{dr} R^m_l &=& - \left( \frac{3}{r}
  - \frac{2 g_0}{c^2} \right) R^m_l
  + \frac{2\mu L_2^2}{r^2} W^m_l
  - \left( 1 - \frac{4\Phi}{c^2} \right) \rho^* \omega^2 W^m_l
  - \frac{4\rho^*}{c^2} ( \stackrel{+}{H}{}^m_l
  + \overline{H}^m_l )  \, . \label{rlm} \nonumber \\
\end{eqnarray}
Up to now, we have fourteen equations (Eqs.(\ref{eq1}),
(\ref{eq2}), (\ref{eq3}), (\ref{eq4}), (\ref{eq5}), (\ref{eq6}),
(\ref{eq7}), (\ref{eq8}), (\ref{ulm}), (\ref{vlm}), (\ref{wlm}),
(\ref{plm}), (\ref{qlm}) and (\ref{rlm})) for the determination of
fourteen unknown functions ($U^l_m$, $V^l_m$, $W^l_m$, $g^l_m$,
$\phi^l_m$, $P^l_m$, $Q^l_m$, $R^l_m$, $B^l_m$, $E^l_m$, $H^l_m$,
$b^l_m$, $e^l_m$, $h^l_m$ ) in contrast to eight equations and
eight unknown functions in the Newtonian limit
(Ref.\cite{wahr82}).


\section{Conclusion}

In this paper we have successfully derived the post-Newtonian
equations for the dynamics of some elastic deformable astronomical
body expanded  in terms of Generalized Spherical Harmonics(GSH).
We have shown that in the Newtonian limit they reduce to the
corresponding Newtonian equations that can be found in Wahr's
paper\cite{wahr82}. The importance to separate a set of partial
differential equation (PDE) into a set of ordinary differential
equation (ODE) is obvious for  numerical calculations. In that
form the equations of motion can be directly used in numerical
computer programs so that the orders of magnitude of the various
relativistic terms for concrete problems can be assessed.

In this paper we considered a non-rotating body as reference state
for our perturbation theory. Since the precessional and nutational
motion of some astronomical body depends upon its dynamical
ellipticity that is related with its rotational motion due to the
inertial forces, this formalism cannot be applied directly to
problems of precession and nutation. The derivation of
corresponding equations and junction conditions represented in
terms of GSH for a rotating elastomechanical ground state will be
the subject of a forthcoming paper.


\acknowledgments

This work was supported by the National Natural Science Foundation
of China (Grant No. 10273008) and the German Science Foundation
(DFG).

\appendix*

\section{Some useful formulae for tensor operations}

Based on the results of Phinney's paper \cite{phin73}, we derive
here some useful formulas for computing the perturbed PN equations.
The expansion of scalars, vectors and tensors is given in
Eqs.~(\ref{gsh1})--(\ref{gsh6}). For simplicity, in the following
we will omit the summation symbols $ \sum_{l=0} ^{\infty} \sum_{m=-l}^l $.

\subsection{Gradient and Laplacian of a scalar $\Phi
(r,\theta,\phi)$}

By means of GSH, a scalar can be expanded as
\[  \Phi (r,\theta,\phi)
 = \phi^m_l(r) D^l_{m0} (\theta ,\phi) \]
and  the gradient is given by
\[ \nabla \Phi = \Phi^{,\alpha} \]
where
\begin{eqnarray}
&& \Phi^{,\pm} =  -\frac{L_1}{r} \phi^m_l D^l_{m\pm1} \\
&& \Phi^{,0} = \frac{d}{dr} \phi^m_l D^l_{m0} \, .
\end{eqnarray}

\noindent
The Laplacian of $\Phi$ can be written as
\begin{equation}
\nabla^2 \Phi = \Phi^{,\alpha\beta}e_{\alpha\beta}
  = \left[ \frac{d^2}{dr^2} + \frac{2}{r}\frac{d}{dr}
  -\frac{l(l+1)}{r^2} \right] \phi^m_l D^l_{m0} \, ,
\end{equation}
where $e_{\alpha\beta}$ are the canonical components of the
identity tensor, its value is given by Eq.(\ref{eab}).

\noindent
For simplicity, we represent $\left[ \nabla (\nabla \Phi ) \right]
^{\alpha\beta} $ as $\Phi ^{, \alpha\beta}$, then the second
partial derivatives of $\Phi$ are given by
\begin{eqnarray}
&& \Phi^{,--} = \frac{1}{r^2}L_1 L_2\phi^m_l D^l_{m-2} \\
&& \Phi^{,-0} = \phi^{,0-}=\left( \frac{L_1}{r^2}\phi^m_l
 -\frac{L_1}{r} \frac{d}{dr}\phi^m_l \right) D^l_{m-1} \\
&& \Phi^{,-+} = \phi^{,+-}=\left( \frac{L_1^2}{r^2}\phi^m_l
 -\frac{1}{r} \frac{d}{dr}\phi^m_l \right) D^l_{m0} \\
&& \Phi^{,00} = \frac{d^2}{dr^2}\phi^m_l D^l_{m0} \\
&& \Phi^{,0+} = \phi^{,+0}=\left( \frac{L_1}{r^2}\phi^m_l
 -\frac{L_1}{r} \frac{d}{dr}\phi^m_l \right) D^l_{m+1} \\
&& \Phi^{,++} = \frac{L_1 L_2}{r^2} \phi^m_l D^l_{m+2}
  \, ,
\end{eqnarray}
where
\begin{eqnarray}
L_1 &\equiv & \sqrt{\frac{l(l+1)}{2}}  \, , \\
L_2  &\equiv & \sqrt{\frac{(l-1)(l+2)}{2}}  \, .
\end{eqnarray}

\subsection{Operations involving a scalar $\Phi (r)$}

If a scalar is a function of the radial coordinate $r$ only, i.e.,
$\Phi = \Phi (r)$,  we have
\begin{eqnarray}
&& \Phi^{,\pm} = 0 \, , \\
&& \Phi^{,0} = \frac{d}{dr} \Phi \, , \\
&& \nabla^2 \Phi = \left( \frac{d^2}{dr^2} + \frac{2}{r}
  \frac{d}{dr} \right) \Phi \, , \\
&& \Phi^{,--} = \Phi^{,++} = \Phi^{,-0} = \Phi^{,0-} =
  \Phi^{,+0} = \Phi^{,0+} = 0 \, , \\
&& \Phi^{,-+} = \phi^{,+-} = - \frac{1}{r} \frac{d}{dr} \Phi \, , \\
&& \Phi^{,00} = \frac{d^2}{dr^2}\Phi \, .
\end{eqnarray}

\subsection{Gradient and divergence of a vector-field}

The expansion  of a vector is given by Eq.(\ref{gsh3}), i.e.,
\begin{equation}
u^\alpha(r,\theta,\phi)= U^{m\alpha}_l(r) D^l_{m\alpha}(\theta\, ,
\phi) \, . \hskip20mm (\alpha=-1 \, , \, 0 \, ,  \, +1)
\end{equation}
We write the vector gradient as
\begin{equation}
u^{\alpha,\beta}(r,\theta,\phi)= U^{m \alpha |\beta }_l(r)
D^l_{m(\alpha + \beta)}(\theta \, , \phi) \, ,
\end{equation}
where
\begin{equation} \label{ap20}
U^{m \alpha |\beta }_l(r) = \left\{ \begin{array}{lll}
 - \frac{1}{r}\left( L_\alpha U^{m\alpha}_l
 + e_{\alpha -} (-1)^\alpha U^{m0}_l
 + \delta_{0\alpha} (-1)^\alpha U^{m-}_l \right)\, ,
 & \beta=-1 \\
 \frac{d}{dr} U^{m\alpha}_l \, , & \beta=0 \\
 - \frac{1}{r}\left( L_{\alpha +1} U^{m\alpha}_l
 + e_{\alpha +} (-1)^\alpha U^{m0}_l
 + \delta_{0\alpha} (-1)^\alpha U^{m+}_l \right)\, ,
 \hskip6mm & \beta=+1
 \end{array}
\right.
\end{equation}
where
\begin{equation}
L_0 = L_1 \, ,
 \hskip16mm
L_{-1} = L_2  \, .
\end{equation}

The symmetric vector gradient can be written as
\begin{eqnarray}
\frac{1}{2} \left( u^{+,-} + u^{-,+} \right)
 &=& \frac{-1}{2r} \left[ L_1 \left( U^{m+}_l + U^{m-}_l
  \right) + 2U^{m0}_l \right] D^l_{m0} \, , \\
\frac{1}{2} \left( u^{\alpha ,0} + u^{0,\alpha} \right)
 &=& \frac{1}{2} \left[ \left( \frac{d}{dr}-\frac{1}{r} \right)
  U^{m\alpha}_l - \frac{L_1}{r}U^{m0}_l \right]
  D^l_{m\alpha} \, .
  \hskip8mm (\alpha=\pm 1)
\end{eqnarray}

The divergence of a vector-fields is expanded in the form
\begin{equation}
\nabla \cdot {\bf u} = u^{\alpha ,\beta} e_{\alpha\beta}
 = \left[ \left( \frac{d}{dr}
 +\frac{2}{r}\right) U^{m0}_l + \frac{1}{r}\sqrt{\frac{l(l+1)}{2}}
 \left( U^{m+}_l + U^{m-}_l \right) \right] D^l_{m0} \, .
\end{equation}

\subsection{Divergence of a symmetric, second order tensor-field}

For a symmetric tensor-field $T^{\alpha\beta}$, the divergence,
($\nabla \cdot \stackrel{\leftrightarrow}{T} )^\alpha
 =T^{\alpha\beta,\gamma}e_{\beta\gamma}=\zeta^\alpha $, is given by
\begin{eqnarray}
\zeta^0 &=& \left[ \left( \frac{d}{dr}
  + \frac{2}{r} \right) T^{m00}_l
 + \frac{2}{r}T^{m+-}_l + \frac{1}{r}\sqrt{\frac{l(l+1)}{2}}
 \left( T^{m0+}_l +T^{m0-}_l \right) \right] D^l_{m0} \, , \\
\zeta^\alpha &=& \left[ \left( \frac{d}{dr} + \frac{3}{r} \right)
 T^{m0\alpha}_l + \frac{1}{r}\sqrt{\frac{l(l+1)}{2}}T^{m+-}_l
 + \frac{1}{r}\sqrt{\frac{(l-1)(l+2)}{2}}
 T^{m\alpha\alpha}_l \right] D^l_{m\alpha} \, .
 \hskip6mm (\alpha=\pm 1) \nonumber \\
\end{eqnarray}

\subsection{Laplacian of a vector-field}

The Laplacian of a vector-field can be written in the form
\begin{equation}\label{nabA}
( \nabla^2 {\bf A} )^\alpha
  = ( A^{\alpha ,\beta ,\gamma} + A^{\beta ,\alpha ,\gamma}
  - A^{\beta ,\gamma , \alpha} )e_{\beta\gamma}
  \equiv \xi ^\alpha \, , \hskip6mm (\alpha=-1, \, 0, \, +1)
\end{equation}
where $A^{\alpha ,\beta}$ is an unsymmetric second order tensor.
In terms of GSH, a tensor-field of second rank can be expanded as
\[
m^{\alpha\beta}(r,\theta,\phi)
  =  M_l^{m \alpha\beta }(r)
  D^l_{m(\alpha +\beta )} (\theta,\phi) \, .
  \hskip6mm (\alpha , \, \beta =-1, \, 0, \, +1)
\]
Differentiation yields
\[ m^{\alpha\beta ,\gamma}(r,\theta,\phi)
  = M_l^{m \alpha\beta | \gamma}(r)
  D^l_{m(\alpha +\beta +\gamma )} (\theta, \phi) \, ,
\]
where the calculation of the quantity  $M_l^{m \alpha\beta |
\gamma}$ is similar to Eq.(2.13) of Ref.\cite{phin73}, note that
$D^l_{mn} = (-1)^{m+n}Y_l^{mn}$. Some terms, used in this paper,
read
\begin{eqnarray}
M_l^{m00|0} &=& \frac{d}{dr} M_l^{m00} \, , \label{ap28} \\
M_l^{m0-|+} &=& - \frac{1}{r} \left( L_1 M_l^{m0-}
  + M_l^{m+-} + M_l^{m00} \right) \, , \label{ap29} \\
M_l^{m0+|-} &=& - \frac{1}{r} \left( L_1 M_l^{m0+}
  + M_l^{m-+} + M_l^{m00} \right) \, , \label{ap30} \\
M_l^{m-0|0} &=& \frac{d}{dr} M_l^{m-0} \, , \label{ap31}\\
M_l^{m--|+} &=& - \frac{1}{r} \left( L_2 M_l^{m--}
  + M_l^{m0-} + M_l^{m-0} \right) \, , \label{ap32} \\
M_l^{m-+|-} &=& - \frac{1}{r} \left( L_1 M_l^{m-+}
  + M_l^{m-0} \right) \, , \label{ap33} \\
M_l^{m+0|0} &=& \frac{d}{dr} M_l^{m+0} \, , \label{ap34} \\
M_l^{m+-|+} &=& - \frac{1}{r} \left( L_1 M_l^{m+-}
  + M_l^{m+0} \right) \, , \label{ap35} \\
M_l^{m++|-} &=& - \frac{1}{r} \left( L_2 M_l^{m++}
  + M_l^{m0+} + M_l^{m+0} \right) \, .\label{ap36}
\end{eqnarray}
Now we have
\begin{eqnarray*}
A^\alpha (r,\theta,\phi)
  &=& A_l^{m \alpha }(r)
  D^l_{m \alpha } (\theta, \phi) \, , \\
A^{\alpha,\beta }(r,\theta,\phi)
  &=& A_l^{m \alpha | \beta }(r)
  D^l_{m(\alpha +\beta )} (\theta, \phi) \, .
\end{eqnarray*}
Set
\begin{equation}
m^{\alpha\beta}(r,\theta,\phi) = A^{\alpha , \beta}
  = M_l^{m \alpha\beta }(r)
  D^l_{m(\alpha +\beta )} (\theta,\phi) \, ,
  \hskip6mm (\alpha , \, \beta =-1, \, 0, \, +1)
\end{equation}
then
\[ (A^{\alpha,\beta})^{,\gamma} (r,\theta,\phi)
 = m^{\alpha\beta,\gamma}(r,\theta,\phi)
 = M_l^{m \alpha\beta | \gamma}(r)D^l_{m(\alpha +\beta +\gamma )}
 (\theta,\phi) \, . \]
The quantities $M^{m\alpha\beta | \gamma}_l (r)$ are derived from
Eqs.(\ref{ap28}), (\ref{ap29}), (\ref{ap30}), (\ref{ap31}),
(\ref{ap32}), (\ref{ap33}), (\ref{ap34}), (\ref{ap35}) and
(\ref{ap36}), and $M^{m\alpha\beta}_l (r) = A_l^{m \alpha | \beta
}(r)$ can be calculated by Eq.(\ref{ap20}), then we can deduce
\begin{eqnarray}
\xi ^0 &=& \left\{ \left[ \frac{d^2}{dr^2} + \frac{2}{r}
\frac{d}{dr}
  - \frac{2}{r^2} (1+L_1^2)\right] A^{m0}_l
  - \frac{2L_1}{r^2} (A^{m+}_l + A^{m-}_l ) \right\} D^l_{m0}
  \, , \label{xi0}  \\
\xi ^- &=& \left\{ \left[ \frac{d^2}{dr^2} + \frac{2}{r}
\frac{d}{dr}
  - \frac{2}{r^2} (1 +L_2^2)\right] A^{m-}_l
  - \frac{2L_1}{r^2} A^{m0}_l  \right\} D^l_{m-1}
  \, , \label{xi-}  \\
\xi ^+ &=& \left\{ \left[ \frac{d^2}{dr^2} + \frac{2}{r}
\frac{d}{dr}
  - \frac{2}{r^2} (1 + L_2^2)\right] A^{m+}_l
  - \frac{2L_1}{r^2} A^{m0}_l  \right\} D^l_{m+1}
  \, .  \label{xi+}
\end{eqnarray}



\end{document}